# Direct Visualization of Local Magnetic Domain Dynamics in a 2D Van der Walls Material/Ferromagnet Interface

Joseph Vimal Vas[1,2#]*, Rohit Medwal[3#], Sourabh Manna[2], Mayank Mishra[2], Aaron Muller[1], John Rex Mohan[4], Yasuhiro Fukuma[4,5], Martial Duchamp[1]*, Rajdeep Singh Rawat[2]*

[1]Laboratory for *in situ* and *operando* Electron Nanoscopy, School of Material Science and Engineering, Nanyang Technological University, Singapore, 9

[2]Natural Sciences and Science Education, National Institute of Education, Nanyang Technological University, Singapore- 637616.

[3]Department of Physics, Indian Institute of Technology, Kanpur, India – 208016.

[4]Department of Physics and Information Technology, Kyushu Institute of Technology, Iizuka 820-8502, Japan.

[5]Research Center for Neuromorphic AI hardware, Kyushu Institute of Technology, Kitakyushu 808-0196, Japan

[#]These authors have contributed equally

*Corresponding authors: rajdeep.rawat@nie.edu.sg, martial.duchamp@gmail.com

## Abstract

Exploring new strategies for controlling the magnetic domain propagation is the key to realize ultrafast, high-density domain wall-based memory and logic devices for next generation computing. These strategies include strain modulation in multiferroic devices, geometric confinement and area-selective pinning of domain wall. 2D Van der Waals materials introduce localized modifications to the interfacial magnetic order, enabling control over the propagation of magnetic domains. Here, using Lorentz-Transmission Electron Microscopy (L-TEM) along with the Modified Transport of Intensity equations (MTIE), we demonstrate controlled domain expansion with *in-situ* magnetic field in a ferromagnet (Permalloy, NiFe) interfacing with a 2D VdW material Graphene (Gr). The Gr/NiFe interface exhibits distinctive domain expansion rate with magnetic field selectively near the interface which is further analysed using micromagnetic simulations. Our findings are crucial for comprehending direct visualization of interface controlled magnetic domain expansion, offering insights for developing future domain wall-based technology.

## Introduction

The propagation of domain walls in magnetic materials, which forms the basis of race track memory[1–3] and domain wall logic circuits[4,5], can be stochastic[6,7]. Controlling the propagation of domain walls with both spin currents and/or external magnetic fields[8] is important for seamless operation of these devices. In order to achieve controlled propagation of magnetic domains, devices based on the localized anisotropy, strain modulation[9–11], geometric confinement[6,12], localized heating[13], external magnetic fields and spin currents[1] have been proposed. These strategies offer unique challenges. For example, in the strain mediated scheme, a local piezoelectric sublayer introduces strain (through lattice distortion) in the magnetic layer by the application of local electric field. This method can be cumbersome to fabricate due to complicated growth requirements for epitaxial piezoelectric layers. The switching speed is also a concern (> 10 ns switching time of ferroelectric domains[14,15]). Similarly, geometric confinement-based domain wall pinning introduces significant challenges in fabrication of the ferromagnet as the dimensions need to be restricted to sub-100 nm[16].

Interaction between 2D Van der Waals (VdW) crystals like graphene or hexagonal boron nitride (h-BN) and ferromagnets (FM) can unlock a new array of fascinating phenomena due to the specific nature of these interfaces. Interfacing graphene with Co induces a giant perpendicular magnetic anisotropy (PMA)[17,18]. Capping a FM with graphene exhibits large interfacial Dzyaloshinskii-Moriya interaction (iDMI) emerged from Rashba-type spin orbit coupling at the graphene/FM interface[19,20]. The iDMI is pivotal for generating, stabilizing and controlling chiral magnetic structures, with potential applications in memory and logic devices. Localized nanoscale iDMI has been demonstrated to effectively achieve domain wall pinning. Notably, the iDMI can be actively tuned through gate voltage modulation, offering electrical control over the chiral magnetic structures and domain wall motion. Therefore, studying the iDMI in the graphene/ferromagnet interface is crucial. Direct visualization using transmission electron microscopy (TEM) can provide a comprehensive understanding of this phenomenon.

The present study experimentally demonstrates a method for controlling the domain wall motion in a ferromagnetic system by introducing local interfaces between ferromagnet and 2D material that exhibit significant iDMI[19]. A 5 nm $Ni_{80}Fe_{20}$ with Gilbert's damping factor of 0.0091 has been chosen as the ferromagnet (see supplementary information section 1 for estimation of the damping factor).

To study the effect of the interface-induced pinning, graphene was transferred on the NiFe thin film. The expansion of the magnetic domains with an external magnetic field near a Gr/NiFe interface is studied to observe the effect of the local iDMI. The magnetic domain mapping require high spatial resolution near the Gr/NiFe interface and thus Lorentz-transmission electron microscopy (L-TEM) is a much-desired tool for such observations[21]. We have performed in-situ domain wall expansion studies using the fringing field of the objective lens of the microscope. The scheme for introducing in-situ magnetic fields for the L-TEM experiments is shown in Fig. 1a. An out of plane magnetic field has been applied on the sample using the magnetic field of the objective lens. The in-plane fields required for the in-situ magnetic field effect studies are introduced by small tilts (along *x* or *y* direction based on alpha or beta tilt) of the sample holder along with the objective lens field which is in the *z* - direction. The geometry of NiFe strips used for the study is shown in Fig. 1b. We have discussed the fabrication of the TEM chips used in this study, in supplementary information section S2. The magnetic field values given in the inset of Fig. 1e are the in-plane magnetic fields for the given objective lens value and tilt conditions.

The initial magnetic domain structure was created by first, saturating the NiFe strips through a small alpha tilt (1.9º) and applying a large objective lens field which is gradually reduced to zero. The normalized $M_x$ and $M_y$ of the NiFe layer, calculated from defocused images (described in methods, Modified Transport of Intensity Equations (MTIE), equations ((3) and (4) are given in Fig. 1d (i) and (ii) respectively. The same area was illuminated throughout the measurements with varying *in-situ* magnetic fields. Distortions were introduced due to the change in focus (objective mini lens) and the objective lens fields used to magnetize the sample. These were compensated by finding a homography transformation matrix[22] to match the distorted image to the in-focus image. A color map of the magnetic signal was then calculated and the corresponding map for Fig. 1d (i-ii) is given in 1e (i). The different domains present in the NiFe can be easily distinguished after the MTIE calculations. The red contrast seen at the edges of the NiFe strip (in Fig. 1e) is the measurement error. This is caused by the improper matching between the in-focus and out-of-focus images due to the Fresnel fringes at the sample edges produced by defocus.

# Results

## In-situ magnetization studies of NiFe

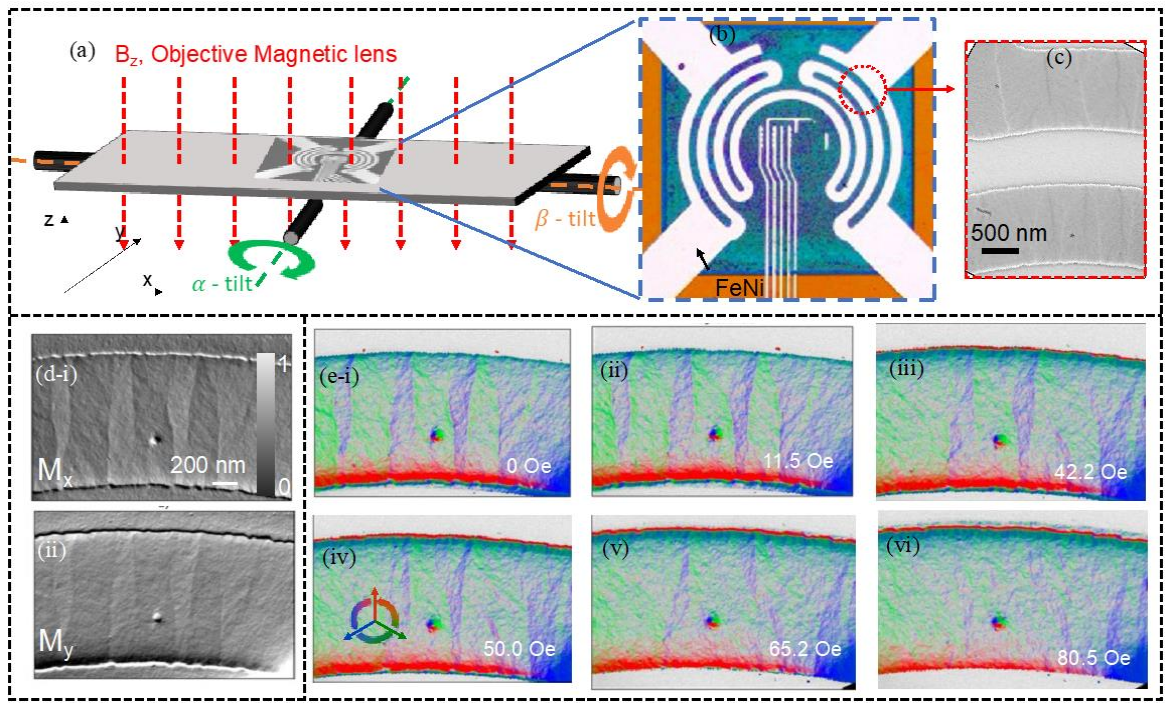

**Figure 1.** *In-situ* TEM experiments on NiFe strips. (a) Schematic of the sample holder and the mechanism for introducing in-plane fields. (b) Geometry of the NiFe strips deposited on the MEMS chip. (c) Out-of-focus TEM image of the NiFe strip showing domain walls. (d) (i) $M_x$ and (ii) $M_y$ of the initial magnetization. (e) Color map of the in-plain domains in the NiFe strips with in-situ magnetic field applied along *x* - direction from (i - vi) 0 – 80.5 Oe.

The spatial evolution of magnetization of NiFe was conducted by applying an external field as shown in Fig. 1e. At 0 Oe, the blue domains (along ~ −x direction) and the green domains (along ~ +x direction) occupy almost equal area on the NiFe strip. The magnetization of the NiFe strip also displays magnetic ripple contrast indicating the polycrystalline nature of NiFe layer[1]. The direction of the electron beam is along z axis as shown in Fig. 1a. The in-situ magnetic field along the *x* - direction was applied by tilting the TEM chip (alpha tilt) by 1.9º and gradually increasing magnetic field by the objective lens field of the TEM using the free lens controller (co-ordinate system and tilt directions with respect to the holder are shown in Fig. 1a). The shrinking of the blue domains with increasing in-plane external field is clearly observed. At 80.5 Oe, the strip is almost completely magnetized along the direction of the green domains (which is the direction of the in-plane field induced by the sample tilt) and the NiFe strip effectively becomes a single domain with in-plane magnetization. The presence of the magnetic ripple contrast arising from local variation in magnetization due to the presence of grains may have increased the saturation field which is comparable with NiFe wires reported in literature[23,24]. The raw images used for the calculation are given in supplementary information, Fig. S5 in section 4.

**Domain wall motion at Gr/NiFe interface**

The next step is to study the domain wall propagation in the vicinity of the Gr/NiFe interface. The preparation of the Gr/NiFe interface is discussed in supplementary section S3. The location of the graphene flake and an image of the Gr/NiFe interface taken in the Lorentz mode are shown in Fig. 2a and 2b. The quality of the Gr/NiFe interface was studied using the cross-section TEM after the experiments were done and is shown in supplementary section S3. The thickness of the graphene was ~ 2 nm indicating about 10 layers. The TEM of graphene showed 6-member carbon rings, (refer Fig. S4 in supplementary information) with the absence of atomistic defects. No amorphous layer was observed between graphene and the polycrystalline NiFe as shown in Fig. S5 of supplementary information indicating a clean VdW/FM interface.

The initial domain wall configuration was produced by introducing both alpha ($\pm 5°$) and beta ($\pm 5°$) tilts, saturating the magnetization using a large objective field and subsequently reducing the magnetic field and the tilts back to 0. The calculated magnetization overlayed onto the in-focus L-TEM images near the Gr/NiFe interface when magnetized along $+x$ and $-x$ directions are given in Fig. 2c and 2d (i-v) respectively. The magnetic field values given in Fig. 2c and 2d are the effective in-plane magnetic fields acting on the sample. The domain walls outside graphene are marked using a white arrow while the domain wall under graphene are marked using a yellow arrow. The defocused LTEM images acquired for different field conditions are given in supplementary information section 4.

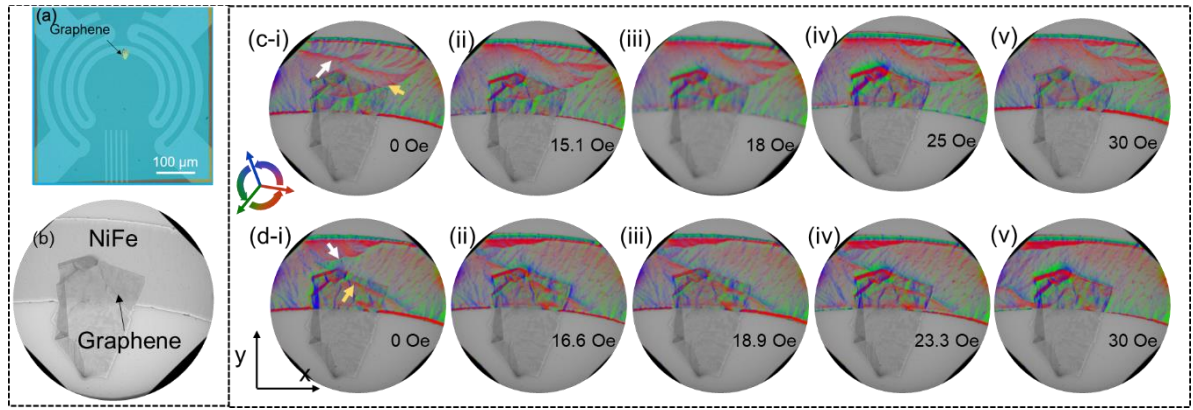

**Figure 2**. (a) Image of the TEM chip showing the location of the Gr/NiFe interface, (b) high resolution TEM image of the Graphene placed on permalloy. Domain wall expansion near the Gr/NiFe interface with in-situ magnetic fields along the (c) (i-v) +x field sweep and (d) (i-v) −x field sweep.

The initial magnetization conditions, characterized by the presence of one domain wall located outside the graphene area and another beneath the graphene, were similar for both field sweeps. The domain walls were subsequently moved by the *in-situ* magnetic fields generated by the objective lens field and the alpha tilt. The field direction was set to move the domain wall to the right (+x direction, 1° alpha tilt) as shown in Fig. 2a(i-v). Increasing the magnetic field to 15 Oe moved the domain wall in uncovered NiFe to the right while the domain wall under graphene remained stationary. Increasing the field even further to 25 Oe, deformed the domain wall under the graphene region slightly but the uncovered domain wall moved further along the NiFe strip. Thus, there is an asymmetry in the motion of the domain walls with magnetic fields under and outside the graphene which is observed for both +x and −x directions. For the $-x$ field sweep, the domain under the graphene switched instantaneously after the in-situ magnetic field of 30 Oe. Similar to the measurements on NiFe strip (Fig. 1e), magnetic ripple contrast from the polycrystalline NiFe grains of varying magnetization have been observed in this case as well.

**Effect of different magnetic energies**

The asymmetry in the motion of the NiFe domain wall under and outside graphene is due to the local modification in the energy density landscape induced by the Gr/NiFe interface. The presence of DMI[19] and PMA[17] at graphene/FM interfaces have been demonstrated using first principle calculations and spin polarized low energy electron microscopy (SPLEEM). Considering these possible interfacial effects, the total magnetic energy density of the NiFe strip in presence of Gr/NiFe interface can be expressed as follows,

$$E_{Total} = E_{Zeeman} + E_{exchange} + E_{magnetostatic} + E_{PMA} + E_{DMI} \qquad (1)$$

In the presence of external magnetic fields, the sizes of the domains are determined in such a way that the total energy of the NiFe strip is minimized. Hence, we carried out micromagnetic simulation in MuMax[3] following the energy minimization routine to understand the possible effect of iDMI and PMA at the Gr/NiFe interface. We model a similar NiFe strip, partially covered with a graphene flake as observed in the L-TEM. The presence of graphene is implicitly modelled by defining a finite DMI and/or PMA only in that particular region which represents the area under the graphene. Detailed methodology of micromagnetic simulation is discussed in the supplementary information, section 5.

We conducted an L-TEM study with a finer in-situ magnetic field step-size of 0.25 Oe to compare the results with micromagnetic simulations. The rate of domain expansion has been quantified by calculating the domain expansion factor, $\epsilon$ which was defined as,

$$\epsilon = \frac{A_{Domain\ 1}}{A_{Domain\ 1} + A_{Domain\ 2}} \quad (2)$$

where $A_{Domain\ 1\ (2)}$ is the area under the domain 1 (2).

The domain expansion factor $\epsilon$ under graphene and outside graphene is given in the Fig. 3(a-i), for each value of the in-situ magnetic field when the domain expands. Note that, both domains are partially covered by the graphene flake. The domain expansion rate as a function of magnetic field, is staggered up to around 25 Oe. This is possibly due to the presence of geometrical defects at the edges which act as the pinning sites. The entire plot is given in supplementary information, section S4. We observe from Fig. 3a, that the slope of domain-expansion rate is a linear function of magnetic field. The rate of domain expansion is smaller at the Gr/NiFe interface (black circles) as compared to its counterpart outside the Gr/NiFe interface (red circle). We have estimated the slopes as 0.0485 Oe$^{-1}$ (from red circles) and 0.0388 Oe$^{-1}$(from black circles) for the area outside and within the Gr/NiFe interface respectively. Hence, we conclude that the presence of graphene indeed slows down or "brakes" the domain expansion in NiFe.

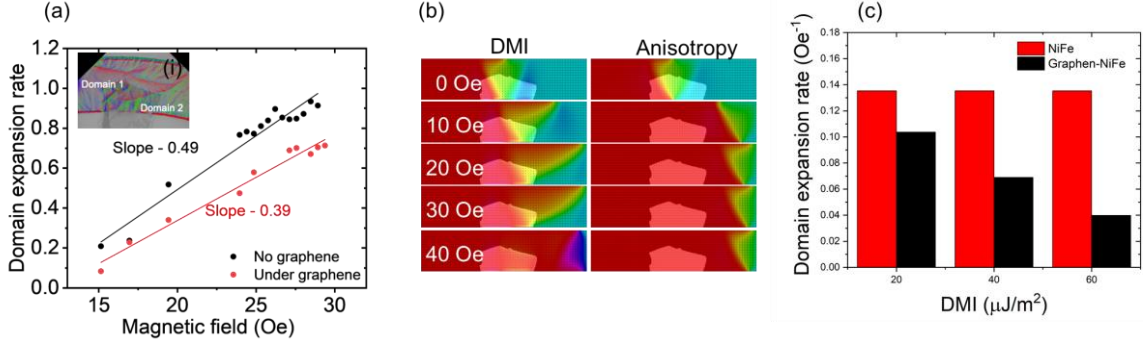

**Figure 3.** (a) Domain expansion rate with magnetic field when the domain expands, (b) simulated expansion of domain within plane magnetic field with anisotropy ($K_u$ = 4.021 kJ/m$^3$) and DMI (2.35 mJ/m$^2$) and (c) simulated rate of domain wall expansion for different DMI values (20 to 60 $\mu$J/ m$^2$).

We have performed micromagnetic simulations to understand the origin of such "braking effect" on domain wall expansion induced by the Gr/NiFe interface. It should be noted that a fully quantitative agreement between the experimental results and simulation could be hardly obtained because of the inevitable mismatch between the simplified micromagnetic model and the real experimental conditions. However, from the qualitative agreement between our experiment and the simulation results, we explain the underlying phenomena. We have first defined a domain wall which is partly Bloch and partly Neel type, at the centre of a rectangular NiFe strip. In addition, we introduced PMA and DMI locally at the Gr/NiFe interface and varied their strength to observe the domain wall motion under the influence of different external magnetic field. A parametric study with PMA has been performed with $K_u$ ranging from 0.4-40.21 kJ/m$^3$ and iDMI strength from 2.3 -2.5 mJ/m$^2$ at the Gr/NiFe interface to observe the possible pinning effect on the domain wall as observed experimentally. This is given in supplementary information, section 5. Softly pinned domain walls were observed with a PMA, $K_u$ = 4.021 kJ/m$^3$ or a iDMI strength of 2.35 mJ/m$^2$ at the Gr/NiFe interface. These are investigated in detail and are shown in Fig. 3b. Even though the domain structure was initialized with a vertical domain wall across the width of the NiFe strip (supplementary information, Fig. S7 (a)), the introduction of PMA and iDMI at the interface changed the shape of the domain wall similar to those observed experimentally. However, the introduction of PMA at the Gr/NiFe interface did not exhibit any pinning effect of domain walls whereas the iDMI at the interface significantly hindered the domain expansion exhibiting the "braking effect". This indicates that the "braking effect" experimentally observed, is because of the interfacial DMI at the Gr/NiFe interface.

The DMI value of 2.35 mJ/m² used in the simulation [shown in Fig. 3(c)] being comparable with values observed by Yang *et al.*[19], appears higher than other experimental measurements[25] on Gr/NiFe interfaces. These experiments report DMI values within the range of 60 μJ/m², with multiple graphene layers reducing the interfacial DMI. This could be because the effect of polycrystalline nature of our NiFe film were not considered in this simulation. To verify this, we carried out micromagnetic simulations with 40 nm grainsize and 10% variation in $M_s$ using the *in-built* Voronoi Tessellation functionality of MuMax[3](which was used to obtain Fig. 3a) with an additional iDMI values in the range of 0- 60 $\mu$ J/m². The domain expansion rate was calculated for different iDMI values and is presented in Fig. 3c. There is a clear reduction in domain expansion rate with increasing iDMI and with an iDMI value of 60 $\mu$ J/ m² the domain expansion rate was comparable to the experimentally observed values. Thus, we conclude that the soft pinning of domain walls observed in *in-situ* L-TEM study near the Gr/NiFe interface was due to iDMI introduced at the interface. This could be used to spatially stifle the expansion of domains to control the domain wall propagation on demand.

## Discussion

We demonstrate that introducing Graphene onto NiFe can reduce the domain expansion rate with magnetic field. Comparison between the Lorentz TEM imaging on NiFe strips with a local Gr/NiFe interface with in-situ magnetic fields with the micromagnetic simulations indicate the reduction in domain wall expansion rate was due to the iDMI introduced by the Gr/NiFe interface. The iDMI strength has been estimated to be around 60 $\mu J/m^2$ based on the simulations considering the variation of dipolar interaction and exchange interaction across the grains to incorporate the microcrystals observed in the L-TEM studies. Spatially arranged interfaces along the length of an FM strip can, therefore, control the expansion of domains which is the key to domain wall memory systems like domain wall logic circuit[5,13,26] and racetrack memory[1,3,27] schemes.

## Methods

### Lorentz Transmission Electron Microscopy (LTEM)

The LTEM imaging was done in a double corrected JEOL ARM 300 microscope in Lorentz mode operated at 300 kV and objective lens turned off. The probe corrector was turned off as well in this mode and the image corrector was tuned to obtain a spatial resolution of ~ 1 nm using a special gold sample of appropriate particle size. For the LTEM imaging, the sample was moved through the focus using the objective mini lens of the image corrector. NiFe devices with and without graphene is prepared on a MEMS chip with a 200 nm thick silicon nitride window which was compatible with a Hennyz TEM holder[21]. The fabrication of the MEMs chips and the transfer of the graphene is discussed in supplementary information S2 and S3.

The transfer protocol was optimized using high resolution transmission electron microscopy to ensure a clean interface between the NiFe and Graphene.

**Modified Transport of Intensity Equations**

The magnetization mapping of the NiFe sample changes the phase of the electron wave passing through, and observed in Lorentz mode when the image is out of focus. The magnetization of the NiFe strip is calculated using the modified transport of intensity equation (MTIE) based on the difference in intensity between the in-focus and out-of-focus images[28]. The equations for the same are given below,

$$\phi(r) = F^{-1}[F(k_z \delta_z I/I)/k_\perp^2] \tag{3}$$

$$tB = \frac{\hbar}{e}[n_z \times \Delta\phi] \tag{4}$$

Where $\phi(r)$ is the phase change in the electron wave introduced by the sample, $F$ and $F^{-1}$ denotes the Fourier and inverse Fourier transform, $k_z$ is the propagation constant of the electron wave in vacuum, $\delta_z I$ represents the change in intensity with defocus, $k_\perp$ is the frequency vector in Fourier space, t is the thickness of the sample, B is the magnetization, $\hbar$ is reduced plank constant, e is the electronic charge and $n_z$ is the unit vector along z direction. Equation 2 has a singularity when $k_\perp$ becomes zero and a low pass filter is used to get rid of this. The electron phase, $\phi(r)$ thus calculated using L-TEM can only provide a relative value for the magnetization of the samples. The code used for the calculation has been uploaded on Github[29].

**Micromagnetic Simulations**

The micromagnetic simulation was performed using the open-source software Mumax3[30] on a rectangular strip of 2048 nm × 512 nm × 5 nm size, discretized into rectangular cells of 2 nm × 2 nm × 5 nm. NiFe material parameters derived from the analysis of FMR spectra were used for the simulation. The exchange constant of NiFe was obtained from the literature as $A_{ex}$ = 13 pJ/m. The Gr/NiFe interface on the NiFe strip was later defined by importing an image mask of the Graphene patch, scaled in accordance with the dimension of the NiFe strip. The spatial distribution of magnetic moments in each cell was initialized as a two-domain configuration where, the magnetization at the left and right halves are defined along the x-axis, facing each other (supplementary information, Fig. S8a). The domain wall (DW) at the middle of the NiFe strip was defined as a mixed state of Bloch wall and Néél wall. DMI and PMA were explicitly defined at the Gr/NiFe interface using the built-in Mumax functions. In each simulation, the final magnetization configuration in the NiFe strip was obtained after relaxing the magnetization to the minimum energy state in presence of external magnetic field, followed by solving the Landau-Lifshitz-Gilbert (LLG) equation. For a detailed discussion on micromagnetic simulation, please see the supplementary information, section 5.

**Acknowledgements**

S.M. acknowledges the support from the NTU-Research Scholarship (NTU-RSS). R.S.R acknowledges the Ministry of Education, Singapore for the support through MOE Tier 2 Grant, ARC-1/17 RSR (MOE2017-T2-2-129) and MOE Tier 1 Grant RG 76/22. Any opinions, findings and conclusions or recommendations expressed in this material are those of the author(s) and do not reflect the views of the Ministry of Education, Singapore.

# Supplementary Information

1) **Ferromagnetic Resonance (FMR) Experiments**

FMR measurements were conducted to study the dynamic properties of the NiFe and the Gr-NiFe interface. The ferromagnetic resonance spectra of FeNi strips and the NiFe thin film with large area graphene were done using the flip chip method. The sample was mounted on a flip-chip coplanar waveguide (CPW) and placed in an in-plane dc magnetic field, $H_{dc}$. This magnetic field, $H_{dc}$, sets up precession within FeNi which damps down in the ns time scale. In-order to set up a continuous precession within FeNi, a radio frequency (rf) magnetic field $H_{rf}$, orthogonal to $H_{dc}$, which is set up by applying an rf current in the frequency range of 14 -20 GHz in the CPW. For a given $H_{dc}$, when the magnetization precession frequency matches the $H_{rf}$, there is sustained oscillations which are sensed using a lock-in based detection scheme. The linewidth $\Delta H$, resonance frequency, $H_0$ were extracted from the FMR spectra by fitting it as the sum of the symmetric and antisymmetric Lorentzian functions,

$$X(H) = L_n \frac{\Delta H (H - H_0)}{(\Delta H)^2 + (H - H_0)^2} + D_n \frac{(\Delta H)^2 - (H - H_0)^2}{(\Delta H)^2 + (H - H_0)^2} \qquad (1).$$

The linewidth, $\Delta H$ vs frequency, f curve and the Kittel fit of the FeNi strips were done to extract properties like Gilbert's damping factor, effective magnetization, $M_{eff}$ and in-plane anisotropy of the material, $H_k$. This is shown in figure S1(ii). The damping factor was extracted by fitting as per the equation $\Delta H = \Delta H_0 + \frac{2\pi\alpha}{\gamma}$. The value of $\gamma$ was kept at 2.9 MHz/Oe. The Kittel fit was done using the equation $f = \sqrt{(H_{dc} + H_k + M_{eff})(H_{dc} + H_k)}$, to calculate the $M_{eff}$ and $H_k$.

The FMR spectra of the NiFe samples in the frequency range of 14- 20 GHz is given in figure S1a (i). The linewidth vs frequency curve used for calculating the Gilbert's damping factor is shown in figure S1a (ii) while the kittel fit is given in figure S1a (iii). The Gilbert damping factor of 0.0091 was observed for the FeNi deposited. The value of $M_{eff}$ was estimated to be 1.017 T.

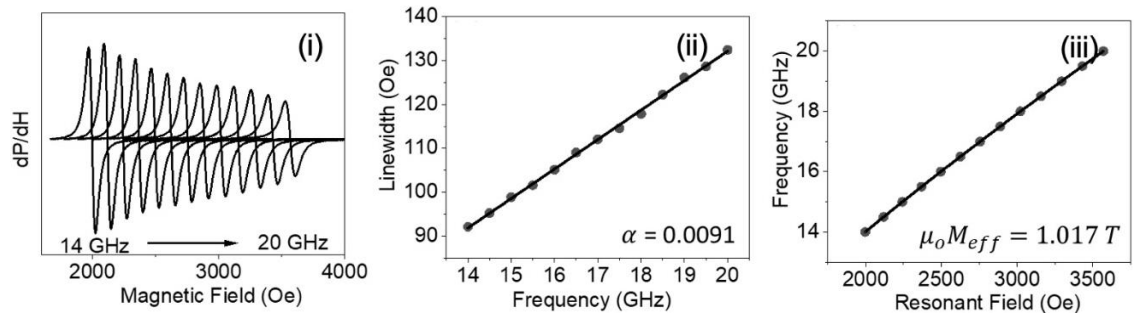

Figure S1. Ferromagnetic resonance spectra of Permalloy.

**2) Chip fabrication for TEM**

Chips compatible with the Hennyz holder meant for JEOL microscopes were fabricated using conventional microfabrication techniques. A silicon wafer with a double side coating of 200 nm LPCVD silicon nitride was used for the fabrication. The fabrication process is pictorially depicted in figure S2. MEMs chips fabricated for such TEM experiments require them to have an electron transparent window on which the material/device to be studied is fabricated. For this purpose, an array of squares is lithographically patterned and reactive ion etched (shown in figure S2, step 2) on the backside before KOH etching 90% of thickness of the silicon (figure S2, step 3). Next, image reversal lithography is used to define the patterns shown in figure S2, step 4 on the frontside of the wafer. The FeNi (5 nm) was deposited on the patterned area using e-beam evaporation technique for minimal impurity and low damping. The metal on top of the remaining resist was then lifted off to obtain the FeNi strips as shown in figure S2, step 6. The wafer was subsequently diced into individual chips which were then used for the Lorentz TEM microscopy.

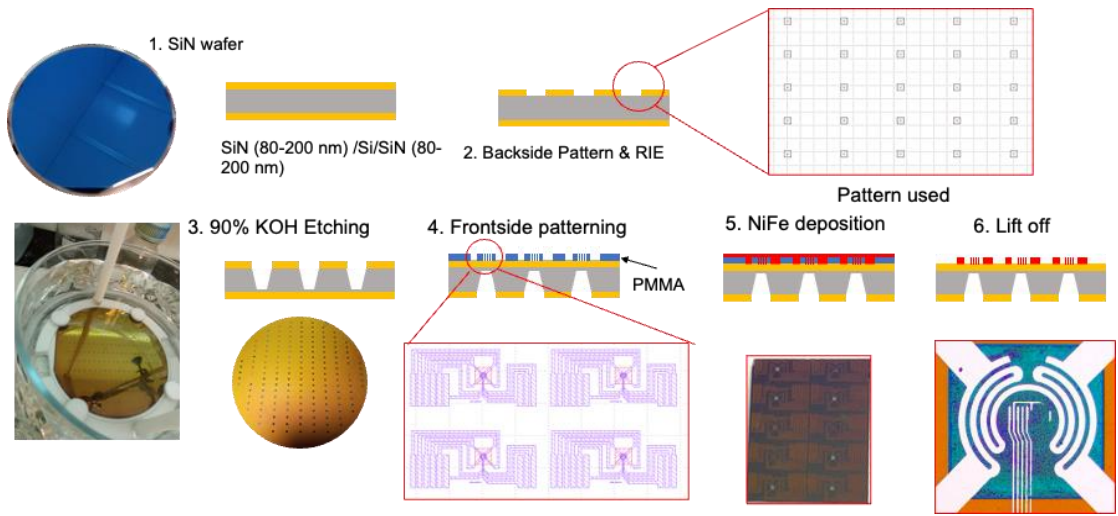

Figure S2. Lithography process for the fabrication of TEM compatible MEMS chips.

### 3) Graphene transfer protocol

The transfer protocol for synthesizing the Gr-FeNi interfaces is given below. Graphene was first transferred onto Nitto tape from a commercial HOPG graphene crystal procured from hq graphene inc. After a couple of fold-unfold self-transfer steps to ensure fewer layers on the tape, the graphene is then transferred onto a PDMS stamp. The protocol for transferring onto the MEMS chip is given in figure S3. The Gr on the PDMS stamp is mounted, facing down, beneath a transparent stage in a confocal microscope. The MEMs chip is mounted onto a heating fixture on a movable stage of the condenser lens holder, which is below the sample stage of the microscope. After locating the graphene flake on the stamp, alignment is done by adjusting the condenser stage x, y, and z locations, as well as the working distance. Once the alignment is done, the stage with the MEMS chip is moved closer to the PDMS stamp with the graphene until the MEMS chip comes in contact with the graphene on the stage. The stage is heated up to 120°C and cooled down until the MEMS chip along with the graphene detaches from the PDMS stamp. The graphene transferred onto the MEMS chip is shown in figure S3 (c).

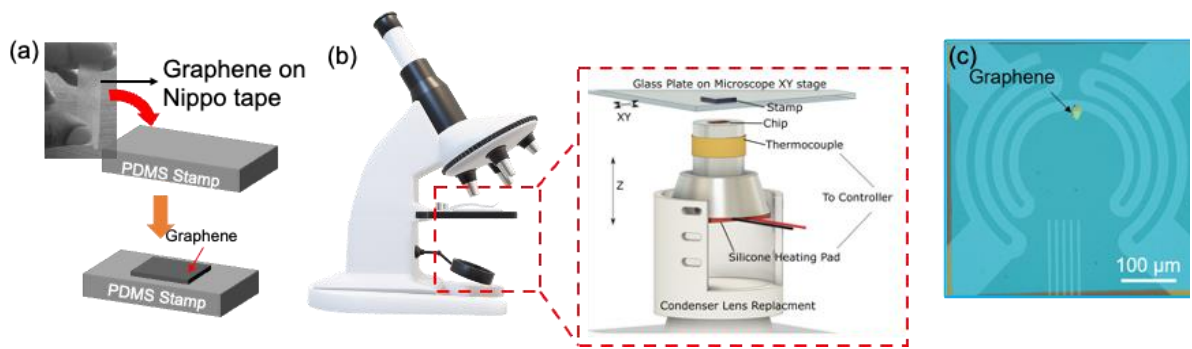

**Figure S3.** Graphene transfer protocol

The quality of graphene transferred was ascertained using TEM imaging. A representative TEM image is given in figure S4. A low magnification image shows a uniform contrast which implies that the graphene transferred is free of large impurities. There are a few patches of amorphous carbon on the surface which were easily removed by plasma cleaning the surface. A representative atomic resolution image of the graphene showing the 6-member carbon rings is shown in the inset of figure S4 which shows the absence of atomistic defects. Rotation of some of the bonds of graphene were observed which could be due to the sheer forces experienced by 2D material during transfer.

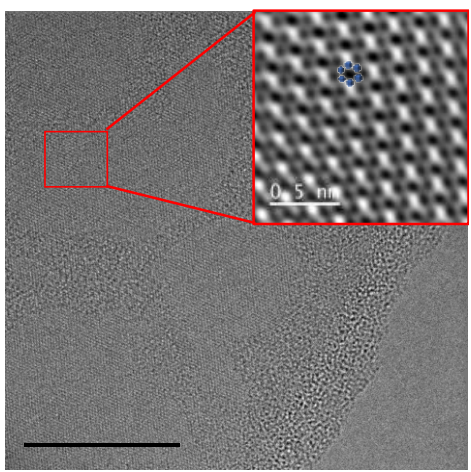

**Figure S4.** TEM images of graphene showing atomic resolution.

The Gr/NiFe interface was studied after the magnetization studies were completed. A cross section of the interface was cut using the focsed ion beam technique and imaged in TEM and Scanning TEM (STEM) mode. This is presented in figure S5. A low magnification high angle annular dark field detector (HAADF) STEM image is given in figure S5(a). A uniform layer of graphene of ~ 2 nm is observed between the e-beam Pt deposited using FIB and the NiFe layer. The thickness of the NiFe layer seems to be ~ 10 nm. The high resolution TEM image of the interface is shown in figure S5(b). 10 – 15 vdW graphene layers can be observed. The polycrystalline NiFe can be seen directly below the vdW Graphene layers without any amorphous layers in between indicating a good interface between the graphene and the NiFe indicating that the amorphous layers seen in figure S4 are above the graphene brought in during exfoliation. The energy dispersive x-ray (EDX) maps of the interface is shown in figure S5 (c).

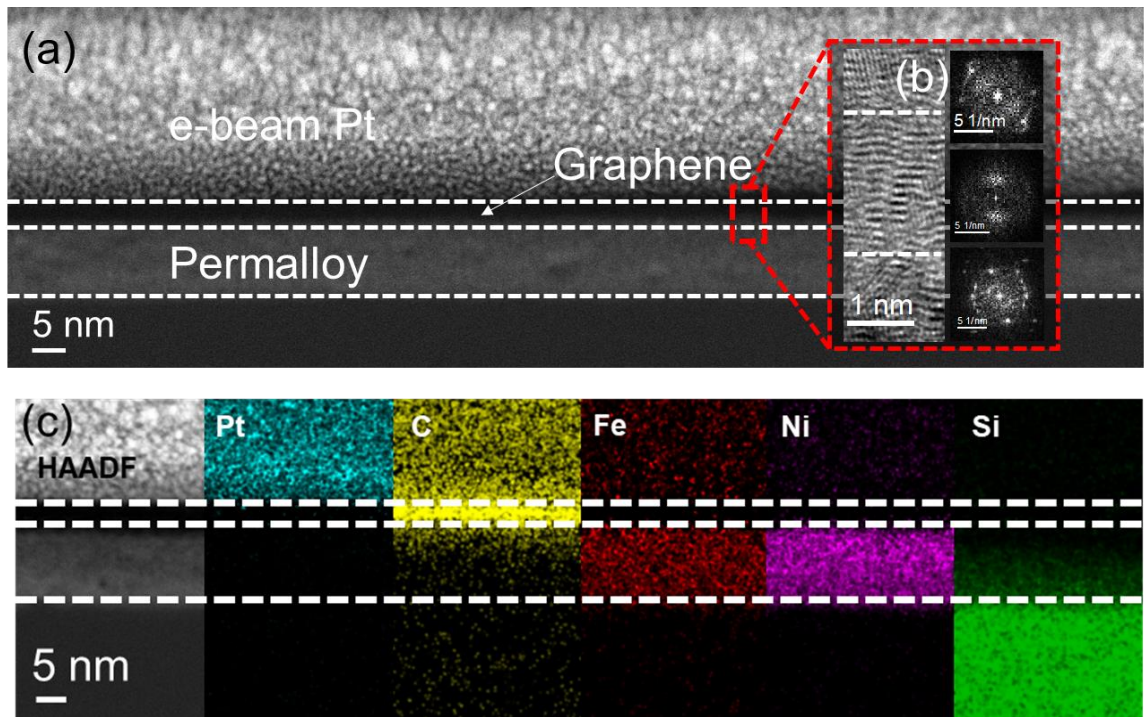

**Figure S5.** Cross sectional TEM images of Gr/NiFe interface. (a) Survey HAADF STEM image of the graphene showing uniform thickness of the graphene, (b) high resolution TEM image showing contact between the vdW layers and the polycrystalline NiFe, (c) EDX maps of the interface showing the location of each of the layers.

4) **Lorentz imaging of Permalloy and Py/Graphene interfaces**

The raw L-TEM images which were used to calculate the magnetization of the domains given in the manuscript are shown below. Figure S5 (a) –(c) shows the in-focus, under-focus and over-focused images of the region of interest of FeNi used for the studies in figure 1. The defocus used for these measurements were 5 mm. The under-focused images of the NiFe strips with different in-situ magnetic fields are given in figure S5 d (i) – (ix). The saturation of the domains can be clearly seen in figure S5 d(ix) where the in plane applied magnetic field was 122 Oe which indicates that the coercive field of the NiFe strips were ~ 100 Oe.

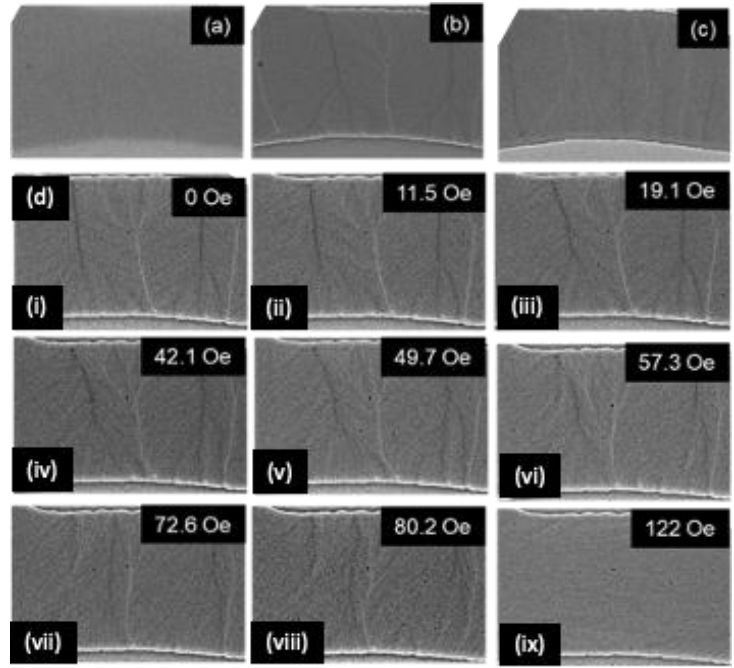

**Figure S5**. L-TEM images of the propagation of domain walls with external magnetic fields

The propagation of domain walls near the Gr-NiFe interfaces are given in figure S6 the magnetic field in the +x direction (figure S6a (i)-(iv)) and in the -x direction (figure S6b (i)-(iv)) . For both +x and -x magnetic field directions, the domain walls outside the Gr-NiFe interface moved easily while a higher field of ~ 25 Oe was required for observing the domain wall propagation under the Gr-NiFe interface. The domains were completely saturated after 30 Oe and the domain wall moved outside the region observed. The magnetization of the regions presented in figure S6b were calculated using the modified transport of intensity equation and presented in figure 2 of the manuscript.

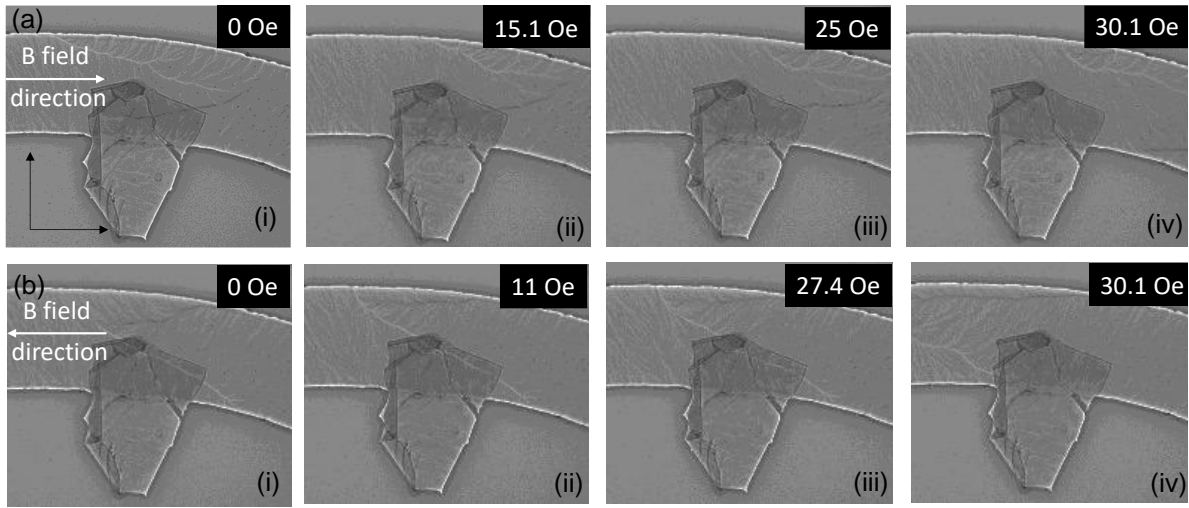

**Figure S6.** DMI induced differentiated domain wall propagation near the permalloy/graphene interface with different magnetic field excitation.

The staggered domain wall expansion with magnetic field is shown in fig. S7. The filled red and black marker dots show this staggered behaviour introduced because of pinning effects from the edges and the polycrystalline NiFe as evidenced by the magnetic ripple contrast. Since the focus of the study is not these effects but the interface effects introduced by the graphene, the domain wall expansion is considered only when the domain was free to move. The selected points when the domain expands is shown as the hollow red and black markers shown in figure S7. The slope of the domain expansion was calculated from these points and shown in figure S7.

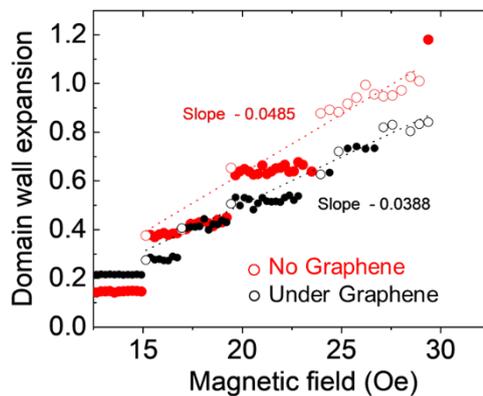

**Figure S7.** Staggered Domain wall expansion with magnetic field

### 5) Micromagnetic simulations

The micromagnetic simulation was performed using the open-source software Mumax3[1] on a rectangular strip of 2048 nm × 512 nm × 5 nm size, discretized into rectangular cells of 2 nm × 2 nm × 5 nm. Different magnetic phenomenon can create the observed Uniaxial shape anisotropy, PMA, iDMI

NiFe material parameters derived from the analysis of FMR spectra (shown in figure S1) were used for the simulation. The exchange constant of NiFe was obtained from the literature as $A_{ex} = 13$ pJ/m. The graphene/NiFe interface on the NiFe strip was later defined by importing an image mask of the Graphene patch, scaled in accordance with the dimension of the NiFe strip. The spatial distribution of magnetic moments in each cell was initialized as a two-domain configuration where, the magnetization at the left and right halves are defined along the x-axis, facing each other (refer to figure S8a). The domain wall (DW) at the middle of the NiFe strip was defined as a mixed state of Bloch wall and Néél wall.

As discussed in the main text, our Graphene/NiFe interface may results in two possible effects− interface induced magnetic anisotropy (iMA) and interfacial Dzyaloshinkii-Moriya interaction (iDMI) [cite]. First, we observed the effect on the domain wall motion by defining a $K_u$ (uniaxial magnetic anisotropy constant) value in the Graphene/NiFe interface using the "region" functionality of Mumax [cite Mumax API/paper] and subsequently relaxing the magnetization to the minimum-energy state in presence of external magnetic field ($B_{ext}$). The final magnetic configurations were obtained after solving the LLG equation in presence of the $B_{ext}$ considering the relaxed magnetic state of NiFe strip as the initial configuration. The easy axis was defined along the (010) axis transverse to the length of the strip. Figure S8b shows the relaxed state of magnetic configuration of the NiFe strip for different values of $K_u$ for $B_{ext} = 0$ (S7b-i to iii, left-side images) and $B_{ext} = 27.8$ Oe (S7b-i to iii, right-side images). Note that $K_u = 4.021$ kJ/m³ shows DW pinning effect (Figure S8-b ii), however, parts of the DW inside and outside the Graphene/NiFe interface move in same speed. In addition, the presence of iMA along (010) axis alters the orientation of magnetic moments in some portion of the Graphene/NiFe interface which is inconsistent with the experimental result. Hence, pure in-plane iMA does not give rise to our experimentally observed DW pinning. Hence, iMA was modified by tilting the easy axis along (011) direction to include interface induced perpendicular magnetic anisotropy (iPMA). Further iDMI was added in presence of iMA by defining the iDMI strength in the Graphene/NiFe interface and observed the final magnetic configuration obtained from the micromagnetic simulation following the same method. We fixed $K_u = 4.021$ kJ/m³. The result of the simulation is presented in figure

S8c-i to iii. It clearly shows DW pinning effect for $D \geq 2.4$ mJ/m$^2$, which is significantly higher iDMI strength for Graphene/NiFe interface. In the next step, the contribution from iMA and iDMI were separated. It is evident that in-plane anisotropy can not explain our experimentally observed DW motion. Hence, in the next step, two different models were simulated separately—(a) DW motion in presence of only iPMA at the Graphene/NiFe interface and (b) DW motion in presence of only iDMI at the Graphene/NiFe interface. The simulation results [refer figure 3b shown in the main text] show that it is the iDMI which influences the DW pinning at smaller field and subsequent saturation at $B_{ext} \geq 30$ Oe, consistent with our experiment. In addition, the motion of DW inside and outside the Graphene/NiFe interface are significantly different—the DW outside the Graphene/NiFe interface moves faster that confirms the pinning at the interface.

In our previous simulations, the polycrystalline nature of the NiFe was not considered. The presence of grains influences the dipolar interaction of the magnetic layer and affects the DW motion. That is a possible reason that the iDMI strength obtained from our simulation result is relatively high. Hence, the NiFe strip was subdivided into grains of average size 40 nm using the Voronoi tessellation[1]. The $M_s$ was randomly varied by 10% across the grains[2] and the inter-grain exchange coupling was reduced by 10%. However, because of the technical limitation of Mumax, the simulation was carried out only within the Graphene/NiFe interface. In addition, the simulation area was approximated to an infinite rectangular strip to eliminate the effects arising from the edges of the graphene patch. The simulation results clearly show the DW pinning due to iDMI where the iDMI strength is consistent with the literature.

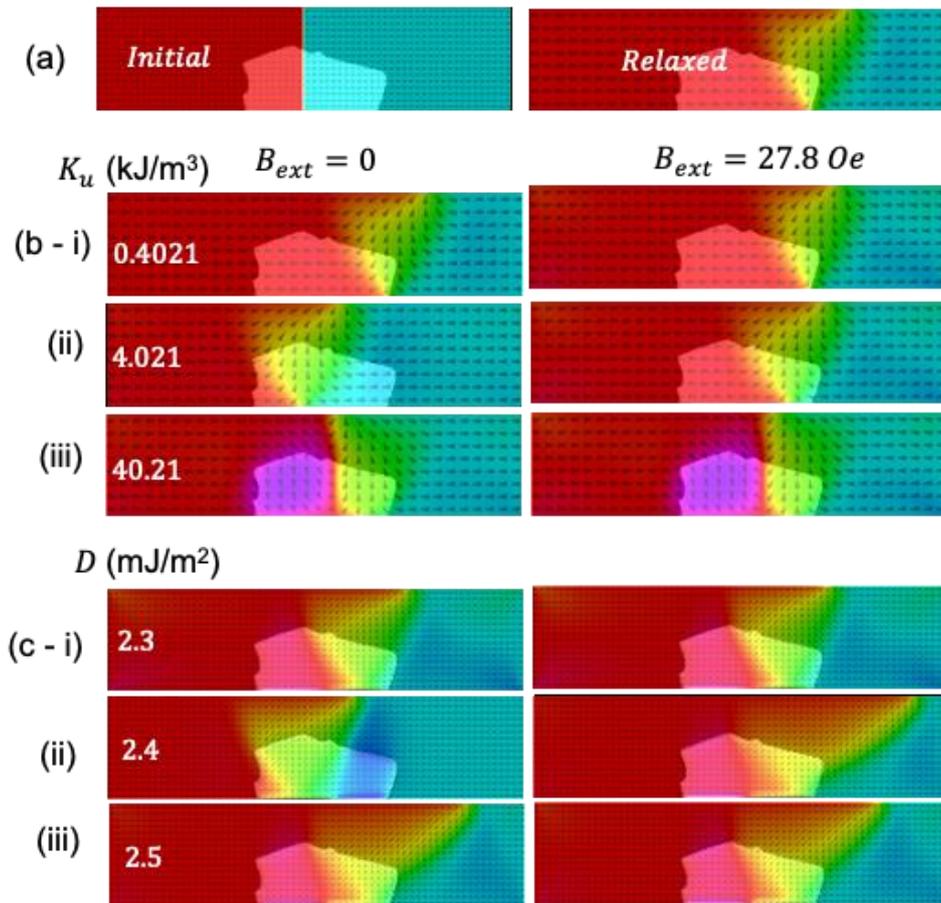

**Figure S8. (a)** Assigned magnetization direction of the strip and the magnetization after relaxiation, **(b)** the propagation of magnetization with magnetic fields for different values of uniaxial anisotropy along (011) direction and **(c)** the propagation of magnetization with magnetic fields for different values of iDMI at the graphene/NiFe interface.

Shape anisotropy,